\documentclass[letter]{aa} 

\newcommand{\Gaia}{{\em Gaia}}

\usepackage{graphicx}
\usepackage{txfonts}

\begin{document}

   \title{The Hyades tidal tails revealed by Gaia DR2}

       \author{Siegfried R\"oser
          \inst{1,2}
          \and
          Elena Schilbach
          \inst{1,2}
          \and
          Bertrand Goldman
          \inst{3,4}
          }
\institute
{Zentrum f\"ur Astronomie der Universit\"at
Heidelberg, Landessternwarte, K\"{o}nigstuhl 12, 69117 Heidelberg, Germany
\and
Zentrum f\"ur Astronomie der Universit\"at
Heidelberg, Astronomisches Rechen-Institut, M\"{o}nchhofstra\ss{}e 12-14, 69120 Heidelberg, Germany\\
    \email{roeser@ari.uni-heidelberg.de, elena@ari.uni-heidelberg.de}
\and
Max-Planck-Institut f\"ur Astronomie, K\"onigstuhl 17, 69117 Heidelberg, Germany
\email{goldman@mpia.de}
\and 
Observatoire astronomique de Strasbourg, Universit\'e de Strasbourg - CNRS UMR 7550,
 11 rue de l'Universit\'e, 67000, Strasbourg, France}

   \date{Received November 7, 2018; accepted ...}

 
  \abstract
   {} 
{Within a 200~pc sphere around the Sun we search for the Hyades tidal tails in the Gaia DR2 dataset.}    
{We use a modified convergent point method to search for stars with space velocities close to the space velocity of the Hyades cluster.} 
{We find a clear indication for the existence of the Hyades tidal tails, a preceding tail extending up to 170~pc from the centre of the Hyades with 292 stars (36 contaminants), and a following tail up to 70~pc with 237 stars (32 contaminants). A comparison with an N-body model of the cluster and its tails shows remarkably good coincidence. Five white dwarfs are found in the tails.}  
   {}


   \maketitle
%

\section{Introduction}
The Milky Way Galaxy exerts tidal forces onto its gravitationally bound
stellar sub-systems with the effect that these sub-systems continuously lose members.
Once the members are no longer gravitationally bound they still could remain in co-moving tidal tails,
preceding and following their home sub-system. The study of the tidal tails of stellar clusters is an important piece of information
on their kinematic evolution, the process of dissolution, and the impact
of the Galactic gravitational field onto a sub-system. Thus far only tidal
tails of massive clusters and dwarf galaxies have been discovered in the Milky Way system. Probably the most famous of these are 
the tidal tails of the globular cluster Palomar 5 (Pal 5) detected by \citet{2001ApJ...548L.165O}. The tails of Pal 5  can be traced over an arc of 10\degr \, on the sky, corresponding to a projected length of 4 kpc at the distance of the cluster \citep{2003AJ....126.2385O}. Tidal tails around open clusters in the Milky Way should also be present, but have not yet been detected. They should be much less prominent than the tails of globular clusters such as Pal 5, as their host clusters, mainly residing close to the Galactic plane with high background density of field stars. Also, possible collisions with
massive molecular clouds can destroy parts of the tails.

The Hyades cluster is the nearest, richly populated open cluster in the Solar neighbourhood. The distance of the centre of the Hyades from the Sun is only  46.75$\pm$0.46 pc  \citep{2017A&A...601A..19G} from Gaia DR1, resp. 47.50$\pm$ 0.15~pc  \citep{2018A&A...616A..10G} from Gaia DR2. This proximity has made it a most interesting target for centuries, and a Simbad query on "Hyades" resulted in 2472 references between 1900 and 2018. \citet{hyades11} found 724 stars co-moving with the mean Hyades space velocity, representing a total mass of 435  M$_\odot$. They determined a tidal radius of about 9 pc, and 364 stars (275 M$_\odot$) are gravitationally bound. \citet{Goldm13} extended this search to lower masses using PanSTARRS1 photometry, down to 0.1\,M$_\odot$. Using the Hipparcos observations \citet{1998A&A...331...81P} determined the isochronic age of the Hyades to be  625$\pm$50 Myr.Using the Hipparcos observations \citet{1998A&A...331...81P} determined the isochronic age of the Hyades to be  625$\pm$50 Myr. Recently, \citet{2018ApJ...856...40M} derived a lithium-depletion-boundary  age of 650 $\pm$ 70~Myr, and \citet{2018ApJ...863...67G} found 680~Myr from MESA stellar evolutionary models including rotation. Using a Bayesian colour-magnitude dating technique including rotating stellar models,  \citet{2015ApJ...807...58B} derived a discordant age for the Hyades of $\approx$800 Myr.

\citet{2009A&A...495..807K} studied the observed and modelled shape parameters (apparent ellipticity and orientation of the ellipse) of 650 Galactic open clusters in the Solar neighbourhood. In the course of this study they performed N-body simulations to follow the evolution of a model open cluster, originally spheroidal, with an initial mass of 1000 M$_\odot$ and the Salpeter IMF down to 0.1 M$_\odot$. This model cluster is moving on a circular orbit in the external tidal
field of the Milky Way and is located close to the observed present location of the Hyades \citep[for more details of the model parameters, see][]{2009A&A...495..807K}. 
It is appropriate to note that the cluster has not been specifically tailored to fit the Hyades, but it is a general model for the evolution of a prototype open cluster in the disk at a Galacto-centric radius of 8.5 kpc.
At a cluster age of 650~Myr, these N-body calculations predicted two tidal tails extending up to ca. 700~pc
from the centre of the cluster along its orbit around the Galactic centre.

In this paper, we report on the detection of the Hyades tidal tails. The paper is structured as follows:
In section \ref{intro} we describe the steps to find the Hyades tidal tails. Section \ref{intro} is divided into subsections describing the astrometric and photometric cleaning, the first detection of the tails after a velocity cut, the cleaning from field stars and the estimation of contamination. The short section \ref{comp} compares observation with the \citet{2009A&A...495..807K} model, and a summary concludes the paper.
\section{Finding the Hyades tidal tails}\label{intro}
According to N-body simulations, a Galactic open cluster, initialised as a spheroid, elongates with time along its
Galacto-centric radius and begins to lose members, mainly low-mass stars which form two tidal tails along the cluster orbit 
\citep{2009A&A...495..807K}. In the N-body models by \citet{2011A&A...536A..64E}, tailored to fit the present-day situation of the Hyades, these tidal tails should
reach a length of about 800 pc if they had not been destroyed in the last 650 Myr by passing molecular clouds, disk shocking, spiral arm
passages and other events not taken into account by the model. Though being sparsely populated, the tails may reveal 
themselves as over-densities of co-moving stars along the cluster orbit.     
With increasing distance from the cluster centre, the space velocities of former cluster members can, however, differ significantly 
from the velocity of the cluster itself. Therefore, we focused our search on a smaller volume and chose a sphere around the Sun with a radius of 200 pc. This should be large enough 
to contain a good portion of the Hyades tails if they still exist.
   \begin{figure}
   \centering
   \includegraphics[width=0.3\textwidth]{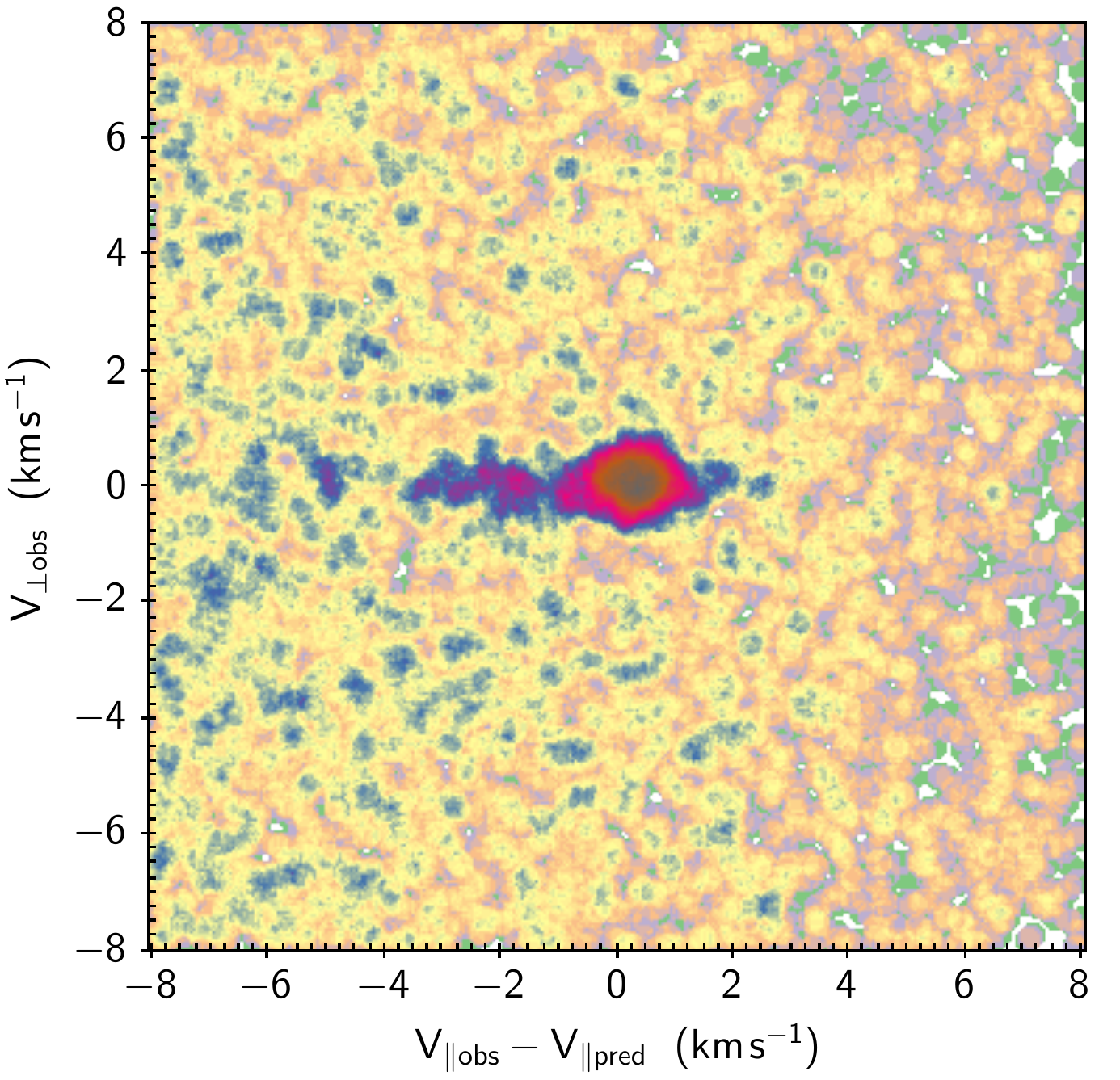}
      \caption{Velocity distribution of the stars in the tangential velocity plane ($V_{\parallel obs}$ -  $V_{\parallel pred}$,  $V_{\bot obs}$). The zero point is the predicted value from the space motion of the Hyades. The x-axes points into the direction of the convergent point.}
         \label{vtan}
   \end{figure}
In the following discussion we use barycentric Galactic Cartesian coordinates $X, Y, Z$. We followed the \Gaia\ catalogue convention and the axes $X, Y, Z$ are directed to the Galactic centre, the direction of Galactic rotation and to the Galactic north pole, respectively. The corresponding velocity coordinates are $U, V, W$. For the central part of the Hyades we determined the phase space coordinates on the basis of positions, proper motions, parallaxes, and radial velocities from Gaia DR2. We found mean values of:
\begin{equation}
\begin{array}{lcl}
\vec{R_c}  =  (X_c,Y_c,Z_c) & = & (-44.77 , +0.40, -16.24)\,{\rm pc}, \\
\vec{V_c}  =  (U_c,V_c,W_c) & = & ( -42.24, -19.00, -1.48)\,{\rm km\,s^{-1}}.\label{COPO}
\end{array}
\end{equation}
These values are almost identical to the previous determination by \citet{2018MNRAS.477.3197R} based on Gaia DR1.
\subsection{Preparatory work}\label{prep}
From the Gaia DR2 dataset \citep{2018A&A...616A...1G} we extracted all
entries having parallax greater than 5 mas, which gave some 6 million objects. 
For the further processing we followed the procedures 
described in \citet{2018A&A...616A...2L}, Chapter 4.3 and Appendix C, Figures C.1 and C.2, to obtain a stellar sample cleaned from possible artefacts.

First we applied the  ``unit weight error"-cut \citep[cf. Eq C.1 in][]{2018A&A...616A...2L} which
removed a considerable portion of dubious entries, and 3.6 million sources remained. 
As a next step we applied the ``flux excess ratio"-cut and followed here the 
procedure given in \citet{2018A&A...616A..10G} which reduced the sample 
to 1,461,162 objects. To exclude dubious measurements we discarded all stars with relative errors of parallax larger than 10 per cent.
This forms an astrometrically and photometrically clean sample of 1,452,246 stars. The distributions of their $G$ and $M_G$ magnitudes show maxima at 17 mag and 12 mag, respectively.
So, the price for this cleaning is some incompleteness at the faint end. 
   \begin{figure}
   \includegraphics[width=0.50\textwidth]{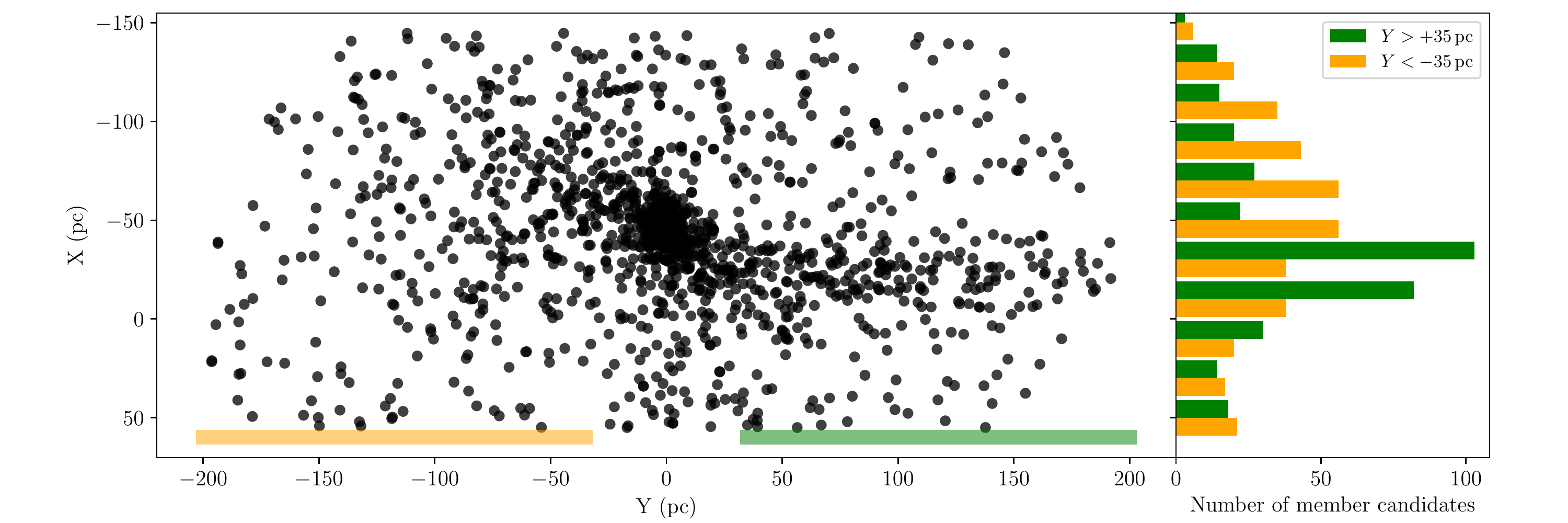}
      \caption{Left: Distribution of the stars in the $Y,X$-plane after the cut in the tangential velocity plane according to the situation in Fig.\ref{vtan}, described in section \ref{CVspace}. Right: Histogram of the marginal distribution in $X$, green: $Y$ > 35~pc, Yellow: $Y$ < -35~pc. }
         \label{tail_raw}
   \end{figure}
%
\subsection{Constraining the velocity space}\label{CVspace}
In the ideal case, we need observed space velocities for each star in a sample to identify stars co-moving with the Hyades. However, measured radial velocities are missing for the vast majority of stars. Therefore we had to rely on criteria based on their tangential velocities, solely. This implies that we may detect highly probable co-moving stars, though they need final confirmation once radial velocities are measured. We followed the formalism of the convergent point (CP) method as described, e.g. in \citet{2009A&A...497..209V} and transformed the Cartesian velocity vector of the cluster motion \vec{V_c} from Eq.\ref{COPO} to give predicted velocities $V_{\parallel pred}$ and $V_{\bot pred}$ parallel and perpendicular to the CP for each star depending on its position on the sky. We note that $V_{\bot pred} \equiv$  0. Also following \citet{2009A&A...497..209V} we similarly transformed the measured (observed) tangential velocities for each star, $\kappa\,\mu_{\alpha*}/\varpi$ and $\kappa\,\mu_{\delta}/\varpi$ into $V_{\parallel obs}$ and $V_{\bot obs}$. Here $\varpi$ is the measured trigonometric parallax in Gaia DR2 and $\kappa=4.74047$ is the transformation factor from 1~mas~yr$^{-1}$ at 1~kpc to 1~km~s$^{-1}$. We also determined the covariance matrix for the velocities $V_{\parallel obs}$ and $V_{\bot obs}$ according to error propagation from
the covariance matrix of the $\mu_{\alpha*},\mu_{\delta}$ and $\varpi$.

To increase the density contrast in the map of the distribution of stars in the tangential velocity plane, we momentarily constrained the volume around the Hyades within we searched 
for the tails. We performed cuts as $|{Z - Z_c}| \leq 20$~pc
and $|{X - X_c}| \leq 100$~pc, where $X_c$ and $Z_c$ are the coordinates of the centre of the Hyades cluster given in
Eq. \ref{COPO}. These cuts are ample compared to the predicted extent of the model tails, and they reduced the sample to 154,389 stars. In Fig.~\ref{vtan} we show the distribution of $V_{\parallel obs}$ -  $V_{\parallel pred}$ and  $V_{\bot obs}$ for volume restricted sample above.
When the space velocity of a star is identical to $\vec{V_c}$ from Eq.\ref{COPO}, 
the differences between the predicted and observed velocities ($V_{\parallel obs}$ -  $V_{\parallel pred}$, $V_{\bot obs}$) must be equal to (0,0).
The strong maximum at (0,0) in Fig.~\ref{vtan} is definitely caused by the
stars of the Hyades cluster proper. Also, there is a clear over-density extending along the x-axis in this plot. To extract the co-moving stars from this over-density, we cut out an area between -5 and +3 km~s$^{-1}$  in $V_{\parallel obs}$ -  $V_{\parallel pred}$ and -0.8 and +1.1 km~s$^{-1}$ in $V_{\bot obs}$ from Fig.~\ref{vtan}. We show the spatial distribution of the 1,580 co-moving stars from these cuts as a projection onto 
the Galactic plane in Fig.~\ref{tail_raw} (left). The most prominent feature is a strong over-density at the position of the Hyades cluster proper at 
$Y, X$ = (+0.40~pc, $-44.77$~pc). The cluster shows an elongated shape indicating that it loses members through the Lagrangian points. Also, we observe a weaker but significant over-density extending from the
Hyades centre towards $Y, X$ = (+190~pc, $-22$~pc) which can be attributed to the preceding tidal tail of the Hyades. 
The beginning of the following tail can be spotted in the direction to $Y, X$ = ($-40$~pc, $-70$~pc) from the cluster centre.
The over-densities related to the preceding and following tails can be clearly seen in Fig.~\ref{tail_raw} (right), where we show the histogram of the marginal distribution in $X$ of stars for Y < -35 pc and Y > 35 pc (avoiding the cluster itself). The peak at X = -30 is due to the preceding tail, and the shallower one at X $\approx$ -55 pc due to the following tail.
However,
there is still a noticeable amount of contamination in Fig.~\ref{tail_raw}. This happens mainly due to a rather generous cut in the tangential 
velocities which allows an increasing number of co-moving field stars to enter the sample. Nevertheless, we need such a large velocity window to identify the tails since the velocities of the tail stars can differ from the Hyades velocity. Stars outside this velocity range are not considered to adhere to the Hyades tails, at least in the context of this paper.
\subsection{Identifying the tidal tails}\label{ITT}
In order to lift out the tails from the background we have to estimate and reduce the influence of isolated field stars that, by chance, obey the velocity restrictions in $V_{\parallel obs}$  and  $V_{\bot obs}$.
In the following, we ignored the volume cuts from the previous section and selected now all stars within the sphere of 200~pc around the Sun that fulfil the velocity restriction (6895 stars). That gave an average stellar density of 0.205$\times 10^{-3}$ stars per pc$^3$. We sub-divided the sphere into cubes with edge lengths of 10~pc to determine the dependence of the contaminating stellar density as a function of the Z coordinate. If all stars would be randomly distributed (which is not the case because, in particular, the Hyades cluster resides in this volume), then we find from Poisson statistics that only 2 of 33,500 cubes should be filled with 4 or more stars. So, to determine the average field star density we disregarded all cubes having 4 or more stars. These presumably contain a non-field star component.
We found a maximum density around $Z$ = -10~pc with a value of 0.358$\times 10^{-3}$ stars per pc$^3$. The density is decreasing for higher and lower $Z$ and falls below the mean of 0.205 at Z = -50~pc and Z = +50~pc. Then, for each cube, we calculated the probability $p(Z)$ that at least one field star from a random distribution is in this cube. For a star in a given cube filled with $N$ stars, we attached an individual value of $p(Z)/N$ as contamination. We did this for all 6895 stars.
This procedure enabled to estimate the number of contaminating field stars in the Hyades cluster itself and its possible tails. This is only a statistical contamination, one cannot spot individual stars as contaminants.

\begin{figure*}
\begin{minipage}[t]{0.3330\textwidth}\vspace{0pt}
\begin{center}
\includegraphics[width=\textwidth]{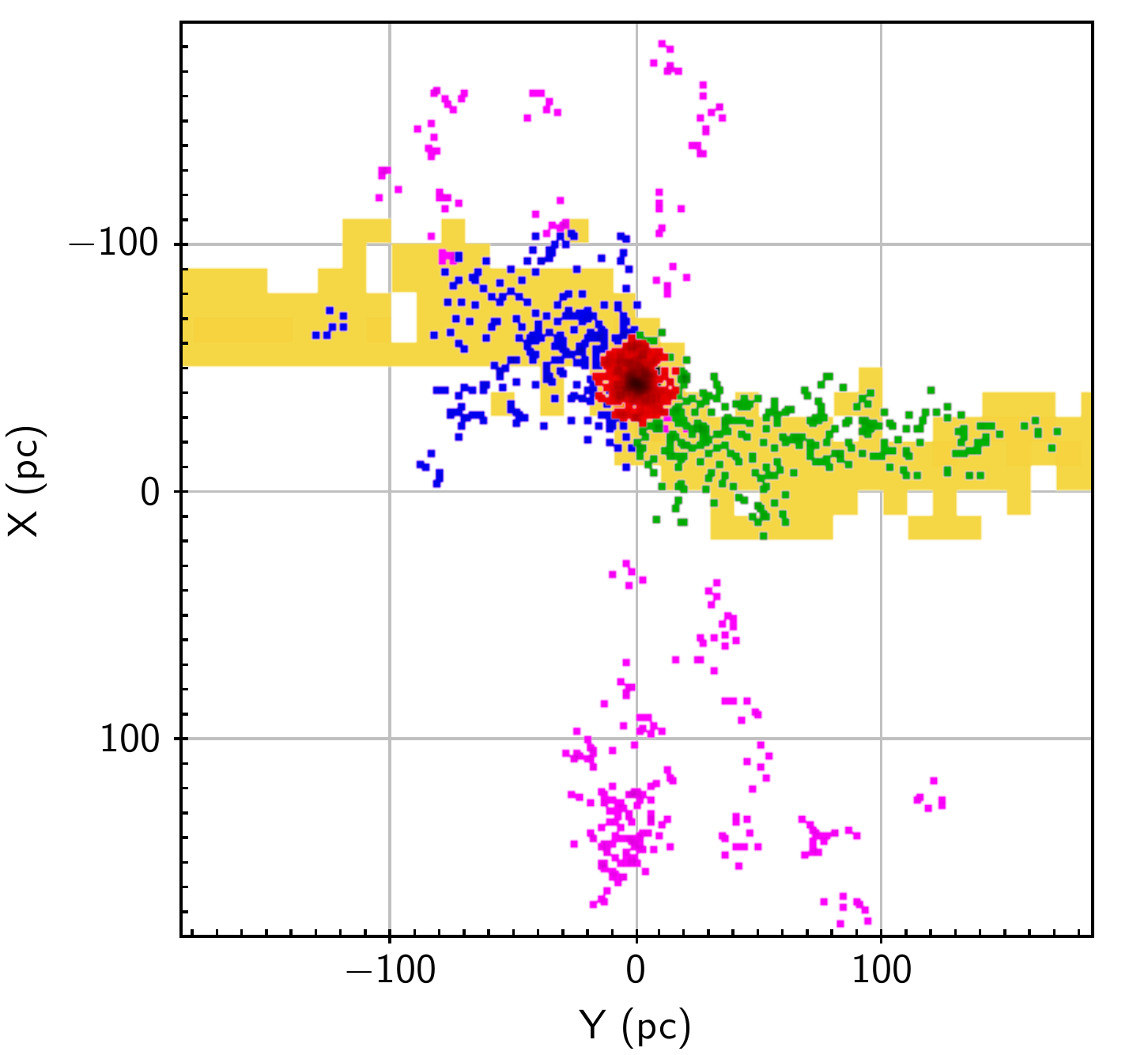}
\end{center}
\end{minipage}\hfill%
\begin{minipage}[t]{0.3330\textwidth}\vspace{0pt}
\includegraphics[width=\textwidth]{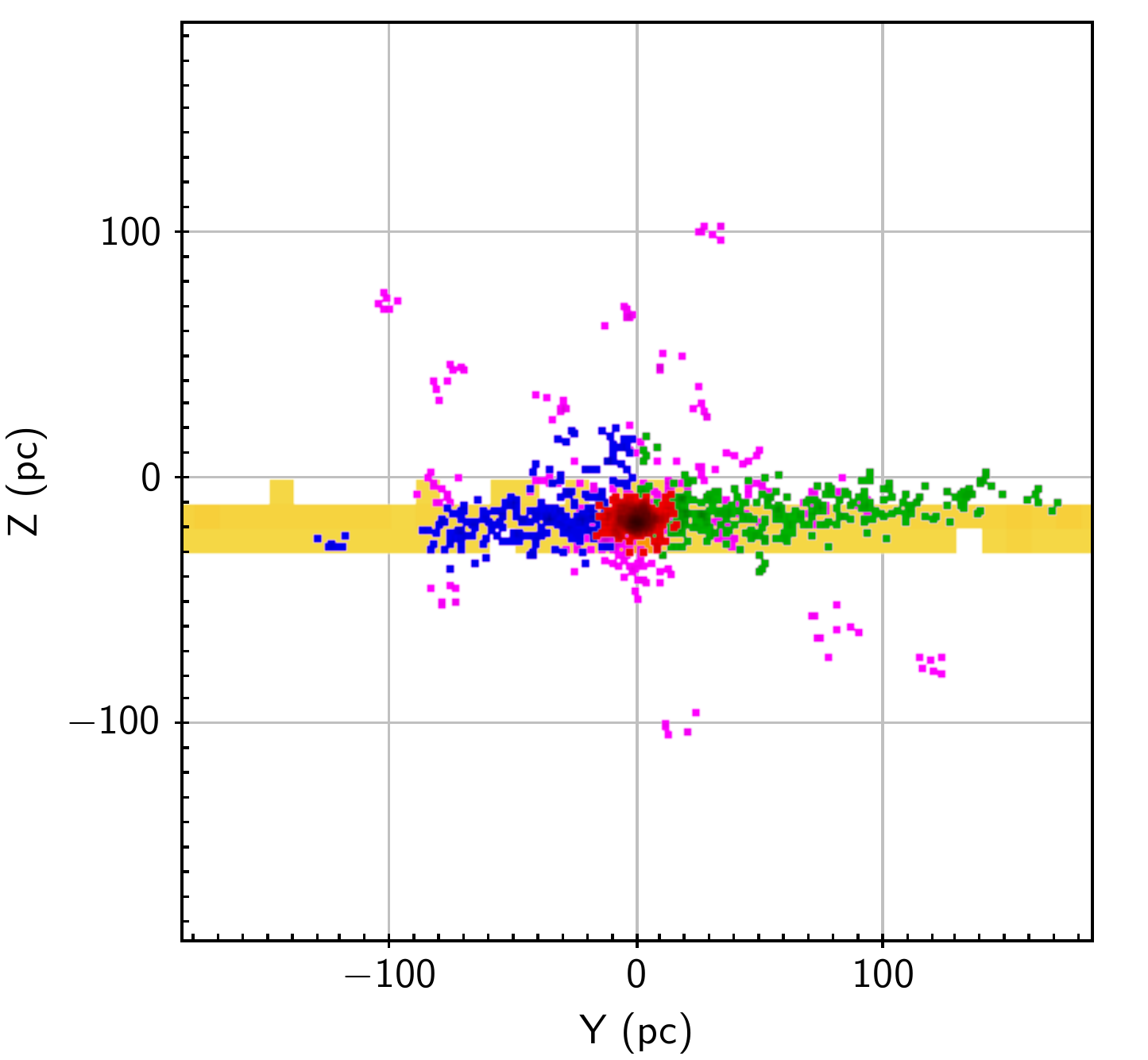}
\end{minipage}\hfill%
\begin{minipage}[t]{0.3330\textwidth}\vspace{0pt}
\includegraphics[width=\textwidth]{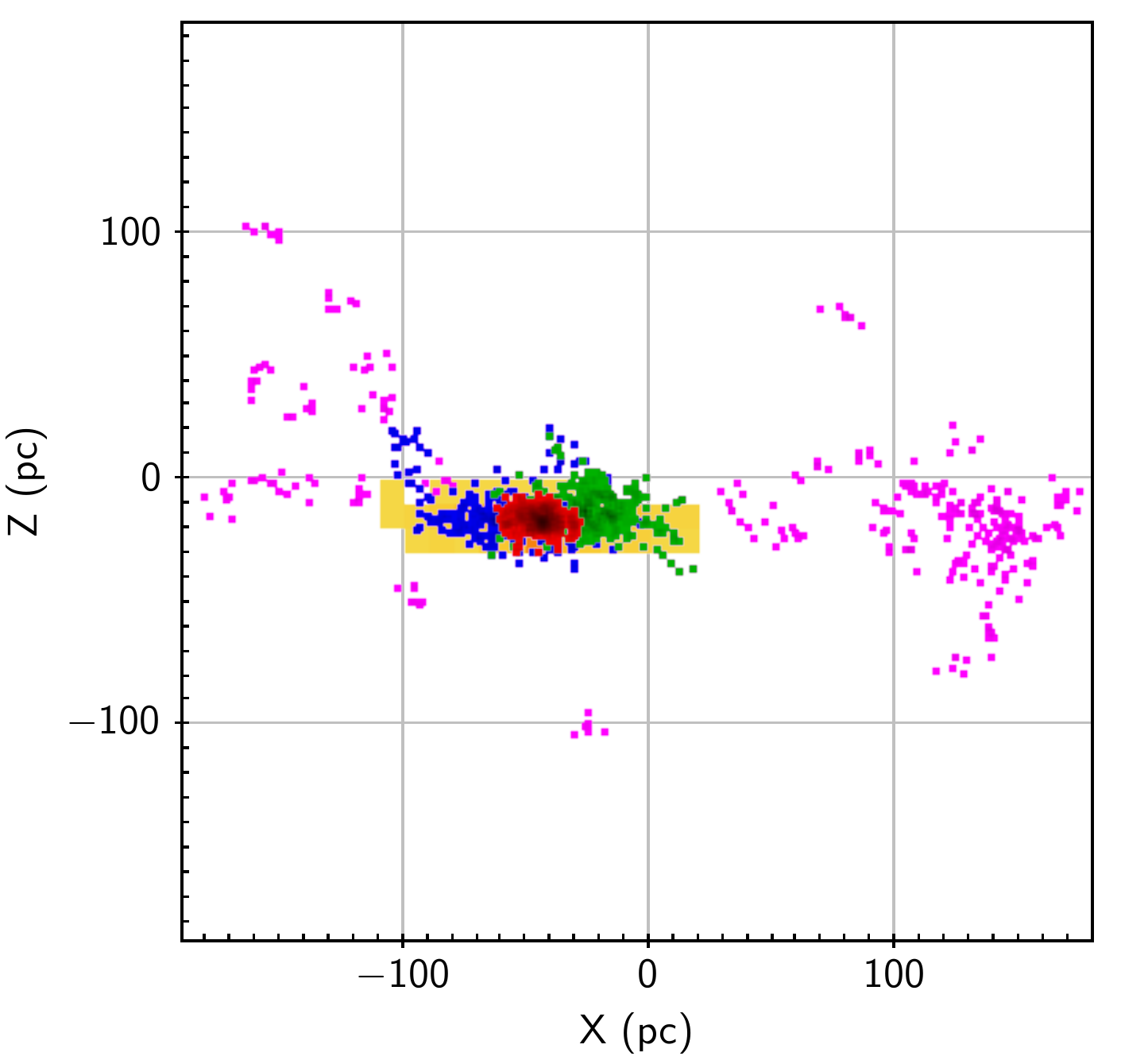}
\end{minipage}\\%
      \caption{The Hyades and their tidal tails. The figures show the spatial distributions of stars in dense regions with at least 2.5$\times 10^{-3}$ stars per pc$^3$ (see text for further explanation of this selection). From left to right: distribution in the $Y,X$-, $Y,Z$-, and $X,Z$-planes. Stars selected in Gaia DR2 are shown as dots, red: the Hyades cluster proper, green: preceding tail, blue: following tail and pink: all other stars in dense regions. The location of the predicted Hyades tidal tails from the model by \citet{2009A&A...495..807K} are indicated by yellow colour in the background.}
\label{3plots}
\end{figure*}

The sub-division of the 200~pc sphere gave rigorous fixed cuts between adjacent cubes irrespective of the actual over-densities. Gradients in a distribution are not well represented, as stars at the edges may fall into an (empty) neighbouring cube. So, to extract physical over-densities we proceeded as follows: around each of the 6895 stars we drew a sphere with radius 10~pc and counted the stars which fell into this sphere. We selected spheres that were filled by 6 stars or more, corresponding to a minimum density of 1.5$\times 10^{-3}$ stars per pc$^3$, and finally selected all stars which belong to at least one of these spheres (1316 stars). The result of this selection is shown in Fig.~\ref{3plots}.
We separated different over-densities of co-moving stars shown by different colours in this figure:
a) The central part of the Hyades defined by a sphere with radius 18~pc around the centre (corresponding to two tidal radii as found by \citet{hyades11}). In this volume we found 501 stars (shown as red dots in Fig. \ref{3plots}). Adding up the contamination of the stars in this volume gives 7. b) Stars in the preceding tail of the Hyades are shown as green dots in Fig~\ref{3plots}. The preceding tail is a rather homogeneous structure extending up to 170~pc in the positive Y-direction We counted 292 stars in this tail with a contamination of 36. c) Blue dots in Fig.~\ref{3plots} mark stars which we assign to the following tail. It looks rather shredded and seems to be in the course of being destroyed after 650~Myr. In the following tail we found 237 stars with a contamination of 32.
As the following tail has such a inhomogenous structure we might have, possibly, chosen some stars that do not originate in the Hyades cluster itself. d) All other stars in this figure are shown as pink dots and represent minor over-densities of co-moving stars falling in the same tangential velocity box as the Hyades itself, but are so far away in space from the Hyades that they probably have nothing to do with the cluster. The most prominent feature is a group of 93 stars centered at $X,Y,Z$ =  (140~pc, -5~pc, -20~pc) in the constellations Sagittarius and Corona Australis, i.e. some 150~pc away from the Sun in the direction to the Galactic centre. A quick look into the CAMD showed that one half of these stars are young (less than 100~Myr) and the other half is of the Hyades age or older.  
\subsection{The Colour-Absolute Magnitude Diagram (CAMD)}\label{CAMDs}
\begin{figure}
   \centering
   \includegraphics[width=0.48\textwidth]{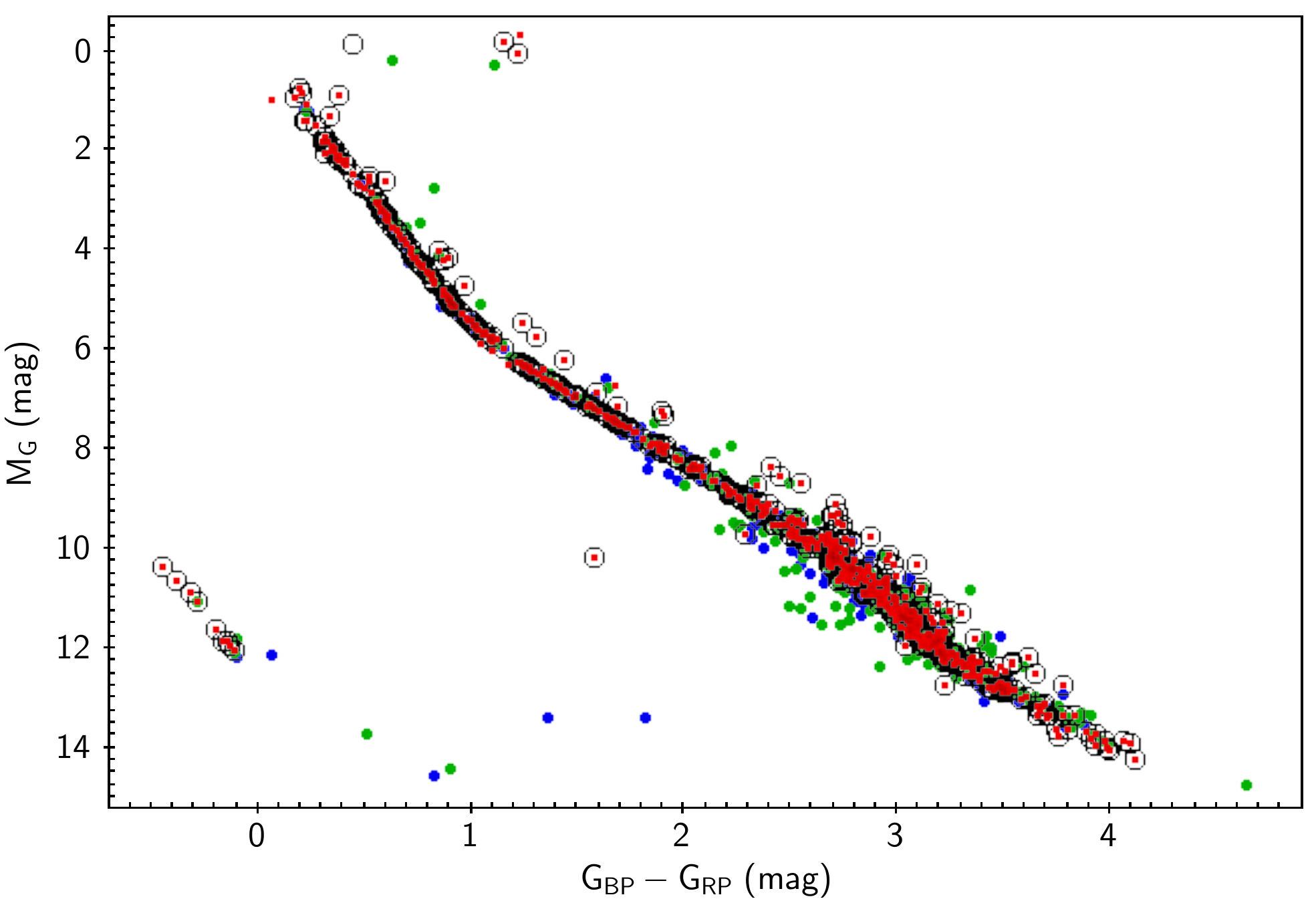}
      \caption{The  $M_G$ vs. $G_{BP} - G_{RP}$ CAMD of the Hyades and their tidal tails. The black circles are the stars by \citet{2018A&A...616A..10G}, representing the cluster proper. The cluster stars selected here are shown as small red dots, and are mostly identical with the Babusiaux stars. The green dots are stars from the preceding tail, the blue dots from the following tail.}
         \label{cmd}
   \end{figure}

\citet{2018A&A...616A..10G} published a sample
of 515 Hyades members, out of which we found 466 in our sample of 1.5 million stars after the Lindegren-cuts. These stars show a sharp and clean empirical cluster sequence in the Colour-Absolute-Magnitude Diagram (CAMD) $M_G$ vs. $G_{BP} - G_{RP}$, which we show in  Figure~\ref{cmd}. In this figure the Babusiaux stars are displayed as black circles, the 501 stars, selected in the previous section as adhering to the cluster itself, are shown as small red dots. The two sequences are identical (small difference in number) and represent the empirical Hyades main sequence. In the case of the tails (blue and green dots) we found some 51 stars below the empirical sequence, i.e. with older ages or different chemistry. These can be regarded as a subset of the 68 contaminants evaluated above, and in these cases we can even identify individual stars as contaminants. Five stars in the CAMD are located at the White Dwarf sequence (M$_G$ < 12.5~mag), 3 in the preceding, 2 in the following tail. Two are already confirmed as white dwarfs. All may have been former Hyades members. 
\section{Comparison between model and observations}\label{comp}

Unfortunately, from the model by \citet{2011A&A...536A..64E}, especially tailored to fit the Hyades cluster as described in \citet{hyades11}, we do not possess a data set representing the simulated tails. However, from the model of \citet{2009A&A...495..807K} we kept a file giving data of the 4000 simulated stars at a time of 650 Myr. Together with the observations in Figure~\ref{3plots} we also show the
distribution of the simulated stars  as a density plot in the background.
We find a remarkable concordance between the shapes of the model distribution and the observations especially with regard to the preceding tail. The following tail in observations is less pronounced in the area with $Y_{gal} < -80$~pc. We can only speculate that it may have been shredded by shocks or collisions with clouds in the last 650 Myr.
%
%
%
\section{Summary}\label{summ}
Using the data from Gaia DR2, we searched for the presence of Hyades tidal tails in a sphere of 200~pc radius around the Sun.
First the Gaia DR2 data were cleaned according to the recipes given in  \citet{2018A&A...616A...2L} and \citet{2018A&A...616A..10G} to obtain an astrometrically and photometrically clean sample. Then, a selection window of only 8~km~s$^{-1}$ by 2~km~s$^{-1}$ in velocity space turned out to be adequate to separate the tails from the vast majority of field stars, at least within our 200~pc sphere.
We found 501 cluster members within two tidal radii (18~pc) from the cluster centre with a possible field star contamination of 7, a preceding tidal tail extending up to 170~pc in the direction of Galactic rotation with 292 stars and 36 contaminants, and a following tail up to 70 pc from the centre of the cluster with 237 stars and 32 contaminants. The latter appears to be shredded. The cluster sequence and the tail sequences in the CAMD are in good coincidence, and the excellent quality of Gaia photometry allowed to actually identify 51 stars below the cluster sequence proper, having older ages and/or different chemistry, which can be considered part of the 69 contaminants. A comparison with a theoretical model for the Hyades tidal tails from N-body calculations \citep{2009A&A...495..807K} showed very good coincidence, especially in the case of the preceding tail, a very satisfactory confirmation for both theory and observations. Five white dwarfs are found in the tails. Serendipitously, we found a moving group, some 150 pc from the Sun near the border between the constellations Sagittarius and Corona Australis. We publish the data of all 1316 stars from Figure~\ref{3plots} as on-line material.
\begin{acknowledgements}
This study was supported by Sonderforschungsbereich SFB 881 ``The Milky Way
System" (subprojects B5 and B7) of the German \emph{Deut\-sche For\-schungs\-ge\-mein\-schaft, DFG\/}. 
This research has made use of the SIMBAD database and of the VizieR catalogue access tool, operated at CDS, Strasbourg, France.
This work has made use of data from the European Space Agency (ESA)
mission \Gaia\ (\url{https://www.cosmos.esa.int/gaia}), processed by
the \Gaia\ Data Processing and Analysis Consortium (DPAC,
\url{https://www.cosmos.esa.int/web/gaia/dpac/consortium}). Funding
for the DPAC has been provided by national institutions, in particular
the institutions participating in the \Gaia\ Multilateral Agreement.
\end{acknowledgements}
\bibliographystyle{aa}
\bibliography{mybib}

\begin{thebibliography}{16}
\expandafter\ifx\csname natexlab\endcsname\relax\def\natexlab#1{#1}\fi

\bibitem[{{Brandt} \& {Huang}(2015)}]{2015ApJ...807...58B}
{Brandt}, T.~D. \& {Huang}, C.~X. 2015, \apj, 807, 58

\bibitem[{{Ernst} {et~al.}(2011){Ernst}, {Just}, {Berczik}, \&
  {Olczak}}]{2011A&A...536A..64E}
{Ernst}, A., {Just}, A., {Berczik}, P., \& {Olczak}, C. 2011, \aap, 536, A64

\bibitem[{{Gaia Collaboration} {et~al.}(2018{\natexlab{a}}){Gaia
  Collaboration}, {Babusiaux}, {van Leeuwen}, {Barstow}, {Jordi}, {Vallenari},
  {Bossini}, {Bressan}, {Cantat-Gaudin}, {van Leeuwen}, \&
  et~al.}]{2018A&A...616A..10G}
{Gaia Collaboration}, {Babusiaux}, C., {van Leeuwen}, F., {et~al.}
  2018{\natexlab{a}}, \aap, 616, A10

\bibitem[{{Gaia Collaboration} {et~al.}(2018{\natexlab{b}}){Gaia
  Collaboration}, {Brown}, {Vallenari}, {Prusti}, {de Bruijne}, {Babusiaux},
  {Bailer-Jones}, {Biermann}, {Evans}, {Eyer}, \& et~al.}]{2018A&A...616A...1G}
{Gaia Collaboration}, {Brown}, A.~G.~A., {Vallenari}, A., {et~al.}
  2018{\natexlab{b}}, \aap, 616, A1

\bibitem[{{Gaia Collaboration} {et~al.}(2017){Gaia Collaboration}, {van
  Leeuwen}, {Vallenari}, {Jordi}, {Lindegren}, {Bastian}, {Prusti}, {de
  Bruijne}, {Brown}, {Babusiaux}, \& et~al.}]{2017A&A...601A..19G}
{Gaia Collaboration}, {van Leeuwen}, F., {Vallenari}, A., {et~al.} 2017, \aap,
  601, A19

\bibitem[{{Goldman} {et~al.}(2013){Goldman}, {R{\"o}ser}, {Schilbach},
  {Magnier}, {Olczak}, {Henning}, {Juri{\'c}}, {Schlafly}, {Chen}, {Platais},
  {Burgett}, {Hodapp}, {Heasley}, {Kudritzki}, {Morgan}, {Price}, {Tonry}, \&
  {Wainscoat}}]{Goldm13}
{Goldman}, B., {R{\"o}ser}, S., {Schilbach}, E., {et~al.} 2013, \aap, 559, A43

\bibitem[{{Gossage} {et~al.}(2018){Gossage}, {Conroy}, {Dotter}, {Choi},
  {Rosenfield}, {Cargile}, \& {Dolphin}}]{2018ApJ...863...67G}
{Gossage}, S., {Conroy}, C., {Dotter}, A., {et~al.} 2018, \apj, 863, 67

\bibitem[{{Kharchenko} {et~al.}(2009){Kharchenko}, {Berczik}, {Petrov},
  {Piskunov}, {R{\"o}ser}, {Schilbach}, \& {Scholz}}]{2009A&A...495..807K}
{Kharchenko}, N.~V., {Berczik}, P., {Petrov}, M.~I., {et~al.} 2009, \aap, 495,
  807

\bibitem[{{Lindegren} {et~al.}(2018){Lindegren}, {Hern{\'a}ndez}, {Bombrun},
  {Klioner}, {Bastian}, {Ramos-Lerate}, {de Torres}, {Steidelm{\"u}ller},
  {Stephenson}, {Hobbs}, {Lammers}, {Biermann}, {Geyer}, {Hilger}, {Michalik},
  {Stampa}, {McMillan}, {Casta{\~n}eda}, {Clotet}, {Comoretto}, {Davidson},
  {Fabricius}, {Gracia}, {Hambly}, {Hutton}, {Mora}, {Portell}, {van Leeuwen},
  {Abbas}, {Abreu}, {Altmann}, {Andrei}, {Anglada}, {Balaguer-N{\'u}{\~n}ez},
  {Barache}, {Becciani}, {Bertone}, {Bianchi}, {Bouquillon}, {Bourda},
  {Br{\"u}semeister}, {Bucciarelli}, {Busonero}, {Buzzi}, {Cancelliere},
  {Carlucci}, {Charlot}, {Cheek}, {Crosta}, {Crowley}, {de Bruijne}, {de
  Felice}, {Drimmel}, {Esquej}, {Fienga}, {Fraile}, {Gai}, {Garralda},
  {Gonz{\'a}lez-Vidal}, {Guerra}, {Hauser}, {Hofmann}, {Holl}, {Jordan},
  {Lattanzi}, {Lenhardt}, {Liao}, {Licata}, {Lister}, {L{\"o}ffler},
  {Marchant}, {Martin-Fleitas}, {Messineo}, {Mignard}, {Morbidelli}, {Poggio},
  {Riva}, {Rowell}, {Salguero}, {Sarasso}, {Sciacca}, {Siddiqui}, {Smart},
  {Spagna}, {Steele}, {Taris}, {Torra}, {van Elteren}, {van Reeven}, \&
  {Vecchiato}}]{2018A&A...616A...2L}
{Lindegren}, L., {Hern{\'a}ndez}, J., {Bombrun}, A., {et~al.} 2018, \aap, 616,
  A2

\bibitem[{{Mart{\'{\i}}n} {et~al.}(2018){Mart{\'{\i}}n}, {Lodieu}, {Pavlenko},
  \& {B{\'e}jar}}]{2018ApJ...856...40M}
{Mart{\'{\i}}n}, E.~L., {Lodieu}, N., {Pavlenko}, Y., \& {B{\'e}jar}, V.~J.~S.
  2018, \apj, 856, 40

\bibitem[{{Odenkirchen} {et~al.}(2003){Odenkirchen}, {Grebel}, {Dehnen}, {Rix},
  {Yanny}, {Newberg}, {Rockosi}, {Mart{\'{\i}}nez-Delgado}, {Brinkmann}, \&
  {Pier}}]{2003AJ....126.2385O}
{Odenkirchen}, M., {Grebel}, E.~K., {Dehnen}, W., {et~al.} 2003, \aj, 126, 2385

\bibitem[{{Odenkirchen} {et~al.}(2001){Odenkirchen}, {Grebel}, {Rockosi},
  {Dehnen}, {Ibata}, {Rix}, {Stolte}, {Wolf}, {Anderson}, {Bahcall},
  {Brinkmann}, {Csabai}, {Hennessy}, {Hindsley}, {Ivezi{\'c}}, {Lupton},
  {Munn}, {Pier}, {Stoughton}, \& {York}}]{2001ApJ...548L.165O}
{Odenkirchen}, M., {Grebel}, E.~K., {Rockosi}, C.~M., {et~al.} 2001, \apjl,
  548, L165

\bibitem[{{Perryman} {et~al.}(1998){Perryman}, {Brown}, {Lebreton}, {Gomez},
  {Turon}, {Cayrel de Strobel}, {Mermilliod}, {Robichon}, {Kovalevsky}, \&
  {Crifo}}]{1998A&A...331...81P}
{Perryman}, M.~A.~C., {Brown}, A.~G.~A., {Lebreton}, Y., {et~al.} 1998, \aap,
  331, 81

\bibitem[{{Reino} {et~al.}(2018){Reino}, {de Bruijne}, {Zari}, {d'Antona}, \&
  {Ventura}}]{2018MNRAS.477.3197R}
{Reino}, S., {de Bruijne}, J., {Zari}, E., {d'Antona}, F., \& {Ventura}, P.
  2018, \mnras, 477, 3197

\bibitem[{{R{\"o}ser} {et~al.}(2011){R{\"o}ser}, {Schilbach}, {Piskunov},
  {Kharchenko}, \& {Scholz}}]{hyades11}
{R{\"o}ser}, S., {Schilbach}, E., {Piskunov}, A.~E., {Kharchenko}, N.~V., \&
  {Scholz}, R.-D. 2011, \aap, 531, A92+

\bibitem[{{van Leeuwen}(2009)}]{2009A&A...497..209V}
{van Leeuwen}, F. 2009, \aap, 497, 209

\end{thebibliography}
\end{document}